\documentclass[amsmath, pra, twocolumn, showpacs]{revtex4-1}
\usepackage{graphicx}

\begin{document}     

\title{Negative azimuthal force of a nanofiber-guided light on a particle}
 
\author{Fam Le Kien}

\author{A. Rauschenbeutel} 

\affiliation{Vienna Center for Quantum Science and Technology, Institute of Atomic and Subatomic Physics, Vienna University of Technology, Stadionallee 2, 1020 Vienna, Austria}

\date{\today}

\begin{abstract}

We calculate the force of a quasicircularly polarized guided light field of a nanofiber
on a dielectric spherical particle. We show that the orbital parts of the axial and azimuthal components of the Poynting vector are always positive while the spin parts can be either positive or negative. 
We find that, for appropriate values of the size parameter of the particle, the azimuthal component of the force is directed oppositely to the circulation direction of the energy flow around the nanofiber. The occurrence of such a negative azimuthal force indicates that the particle undergoes a negative torque. 
\end{abstract}

\maketitle

\section{Introduction}
\label{sec:introduction}

It is well known that light has not only energy but also linear and angular momenta that can be transferred to atoms, molecules, and material particles. Radiation pressure due to the momentum flux in a light beam tends to push illuminated objects along the direction of propagation. The intensity-gradient force tends to draw small objects toward extrema of the intensity.
In single-beam optical traps known as optical tweezer, the axial intensity gradient is deep enough that the intensity-gradient force exceeds radiation pressure \cite{Ashkin86}. The tightly focused light beam of an optical tweezer can therefore draw particles against the light propagation direction. Like the optical tweezer technique \cite{Ashkin86,Fazal11,Padgett11,Dholakia11,Juan11}, the techniques of optical solenoid beams \cite{Lee10} and optical conveyor belts \cite{Cizmar05,Cizmar06,Ruffner12} also require gradient forces to move small objects back and forth along the beam axis.

Recent studies have shown theoretically \cite{Sukhov10,Chen11,Novitsky11,Novitsky12,Sukhov11,Saenz11} and demonstrated experimentally \cite{Brzobohaty13} that small particles can be pulled by so-called tractor light beams against the photon stream even when the beam
intensity is uniform along the propagation axis. Unlike other optical techniques of micro-object manipulation \cite{Ashkin86,Lee10,Fazal11,Padgett11,Dholakia11,Juan11,Cizmar05,Cizmar06,Ruffner12}, the pulling of particles against the propagation direction of a tractor beam originates from a non-conservative optical backward pulling force. Such a  pulling force occurs when
the projection of the total incident photon momentum along the propagation direction is small and the forward scattering is dominant \cite{Chen11,Novitsky11,Novitsky12,Sukhov11,Saenz11,Brzobohaty13}.

Several types of tractor beams have been proposed and studied. They include interfering Bessel beams \cite{Sukhov10}, interfering plane waves \cite{Sukhov11,Saenz11,Brzobohaty13}, or a single Bessel beam with a semi-apex angle close to $90^\circ$ \cite{Chen11,Novitsky11,Novitsky12}. A disadvantage of the Bessel-based technique is that, as with a plane wave, a true Bessel beam cannot be created because it is unbounded and would require an infinite amount of energy.
Approximations to Bessel beams are made in practice by focusing a Gaussian beam with an axicon lens to generate a Bessel-Gauss beam.
Such beams exist only in a spatially limited volume dictated by optical elements. Consequently, 
the lengths of approximate versions of Bessel beams are short (typically less than 10~$\mu$m). Similarly, the lengths of tractor beams created by interfering Gaussian beams, which are approximations to plane waves in practice, are also limited \cite{Brzobohaty13}.

Due to wide applications of near-field optical microscopy and spectroscopy, especially in biology, medicine, materials  engineering, and information technology, a lot of attention has been paid to scattering of evanescent waves by small particles.
The motion of small particles in the evanescent field produced by total internal reflection of a laser beam from a dielectric surface has been experimentally studied \cite{Kawata92}. The first theory for scattering of evanescent waves by a spherical particle, developed by Chew \textit{et al.} \cite{Chew79} and slightly corrected by Liu \textit{et al.} \cite{Liu95}, is essentially the analytical continuation of the standard Mie theory for the case of plane-wave excitation \cite{Mie books}. The generalized Lorentz-Mie theory for a spherical particle in an arbitrary incident light field has been developed  \cite{Barton88,Gouesbet88,Barton89,GLMT,Salandrino12}. Differential cross sections \cite{Liu95} and total cross sections \cite{Quinten99} for extinction of evanescent waves by small spherical particles have been calculated.
The radiation force exerted on small spherical particles in an evanescent field near a dielectric interface has been studied \cite{Prieve93,Almaas95,Lester99,Chaumet00,Brevik03,Nieto-Vesperinas04,Siler06,Skelton12,Almaas13}.

An interesting scheme for generating an evanescent-wave optical field is to use an optical fiber that is tapered to a diameter     comparable to or smaller than the wavelength of light \cite{Mazur's Nature,Birks,taper,Morrissey13}. In such a thin fiber, called a nanofiber, the guided field penetrates an appreciable distance into the surrounding medium and appears as an evanescent wave carrying a significant fraction of the propagating power and having a complex polarization pattern \cite{Bures99,Tong04,fibermode}.
When the fiber diameter is not too small, the guided field is tightly confined in the fiber transverse plane \cite{Bures99,Tong04,fibermode}. Nanofiber-guided light fields can be used for trapping atoms \cite{fiber trap,Vetsch10,Goban12}, for probing atoms \cite{absorption,Nayak07,Nayak09,Dawkins11,Reitz13,Russell13}, molecules \cite{Stiebeiner09}, quantum dots \cite{Yalla12}, and nanodiamonds \cite{Schroder12,Liebermeister13}, and for mechanical manipulations of small particles \cite{Skelton12,Brambilla07}. Recently, it has been demonstrated that small particles can be attracted to and transported along nanofibers by the evanescent wave of a guided light beam \cite{Skelton12}.

In this paper, we study the force of the evanescent wave of a quasicircularly polarized guided light field of a nanofiber
on a dielectric spherical particle. We show that, for appropriate values of the size parameter of the particle, the azimuthal component of the force is directed oppositely to the circulation direction of the energy flow around the nanofiber. 

The paper is organized as follows. In Sec.~\ref{sec:nanofiber} we describe the nanofiber and study the Poynting vector of 
a quasicircularly polarized guided light field. In Sec.~\ref{sec:force} we investigate the optical force of the nanofiber-guided light field on a dielectric particle. 
Our conclusions are given in Sec.~\ref{sec:summary}.

\section{Quasicircularly polarized nanofiber-guided light field}
\label{sec:nanofiber}

\begin{figure}[tbh]
\begin{center}
  \includegraphics{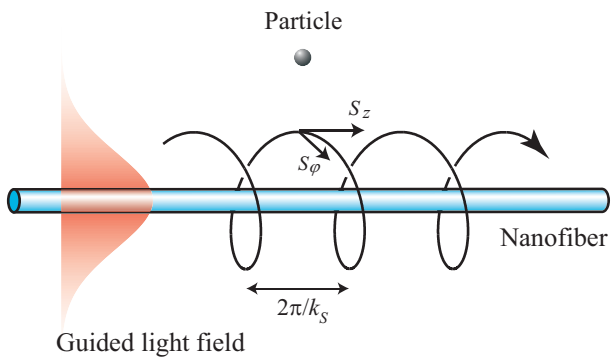}
 \end{center}
\caption{(Color online) A dielectric particle in the vicinity of a nanofiber.
The components and trajectory of the Poynting vector of the field in a
quasicircularly polarized fundamental mode are shown. The period of the trajectory is $2\pi/k_S$ where $k_S=S_{\varphi}/rS_z$.}
\label{fig1}
\end{figure}

We consider the mechanical action of a nanofiber-guided light field on a dielectric spherical particle (see Fig.~\ref{fig1}). The nanofiber is a silica cylinder of radius $a$ and refractive index $n_1$ and is surrounded by an infinite background medium of refractive index $n_2$, where $n_2<n_1$. 
We neglect the absorption by the nanofiber material and the surrounding medium, that is, we assume that $n_1$ and $n_2$ are real parameters.
The radius $a$ of the nanofiber is well below the wavelength $\lambda$ of light. Therefore, the nanofiber supports only the hybrid fundamental modes HE$_{11}$ corresponding to the given wavelength \cite{fiber books}.

In order to describe guided light fields, we use Cartesian coordinates  $\{x,y,z\}$ and associated cylindrical coordinates $\{r,\varphi,z\}$, with $z$ being the fiber axis. We represent the electric and magnetic components of the field as
$\mathbf{E}=(\boldsymbol{\mathcal{E}}e^{-i\omega t}
+\boldsymbol{\mathcal{E}}^\ast e^{i\omega t})/2$ 
and $\mathbf{H}=(\boldsymbol{\mathcal{H}}e^{-i\omega t}
+\boldsymbol{\mathcal{H}}^\ast e^{i\omega t})/2$, respectively. Here, $\omega$ is the angular frequency of the field, and $\boldsymbol{\mathcal{E}}$ 
and $\boldsymbol{\mathcal{H}}$ are the complex amplitudes of the electric and magnetic components, respectively.
We study the case where the light field is quasicircularly polarized.
We assume that the principal rotation direction of the field polarization
around the fiber axis $z$ is counterclockwise and the light field propagates in the positive direction of the axis $z$. 
In this case, the amplitudes of the field are \cite{fiber books,fibermode}
\begin{equation}
\boldsymbol{\mathcal{E}} 
= \mathcal{N}(\hat{\mathbf{r}}e_r+\hat{\boldsymbol{\varphi}}e_\varphi+
\hat{\mathbf{z}}e_z) e^{i\beta z +i\varphi}
\label{p9}
\end{equation}
and 
\begin{equation}
\boldsymbol{\mathcal{H}} 
= \mathcal{N}(\hat{\mathbf{r}}h_r+\hat{\boldsymbol{\varphi}}h_\varphi+
\hat{\mathbf{z}}h_z) e^{i\beta z +i\varphi}.
\label{p10}
\end{equation}
The propagation constant $\beta$ is determined by the fiber eigenvalue equation (see \cite{fiber books} and Appendix \ref{sec:mode}). 
The cylindrical components $\{e_r, e_\varphi, e_z\}$ and $\{h_r, h_\varphi, h_z\}$ of the mode-profile vector functions $\mathbf{e}(\mathbf{r})$ and $\mathbf{h}(\mathbf{r})$ of the electric and magnetic parts, respectively, of the fundamental guided modes are given in Appendix \ref{sec:mode}.
The coefficient $\mathcal{N}$ is determined from the power $P_z$ of the light field.

\begin{figure}[tbh]
\begin{center}
  \includegraphics{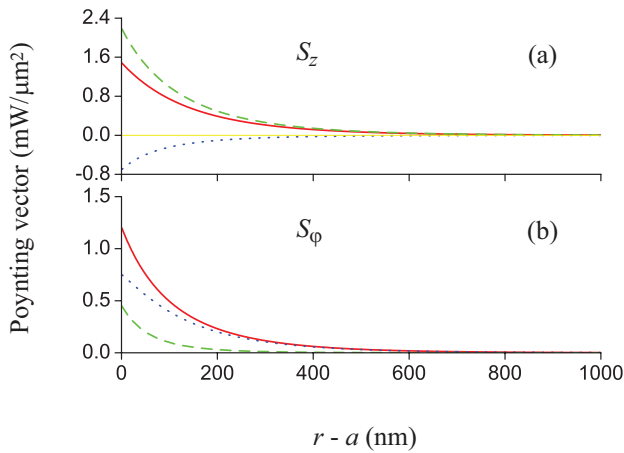}
 \end{center}
\caption{(Color online) Radial dependencies of the axial component $S_z$ (a) and the azimuthal component $S_\varphi$ (b) of the Poynting vector of the guided light field in a counterclockwise quasicircularly polarized fundamental mode. The total values are shown by the solid red curves while 
the orbital and spin parts are shown by the dashed green and dotted blue lines, respectively.
The wavelength and the power of light are $\lambda=1064$ nm and $P_z=1$ mW, respectively. 
The radius and refractive index of the nanofiber are $a=250$ nm and $n_1=1.45$, respectively.
The medium surrounding the nanofiber is vacuum, with the refractive index $n_2=1$.
The yellow line in part (a) stands for the zero value of the Poynting vector and is a guide to the eye.
}
\label{fig2}
\end{figure}

\begin{figure}[tbh]
\begin{center}
  \includegraphics{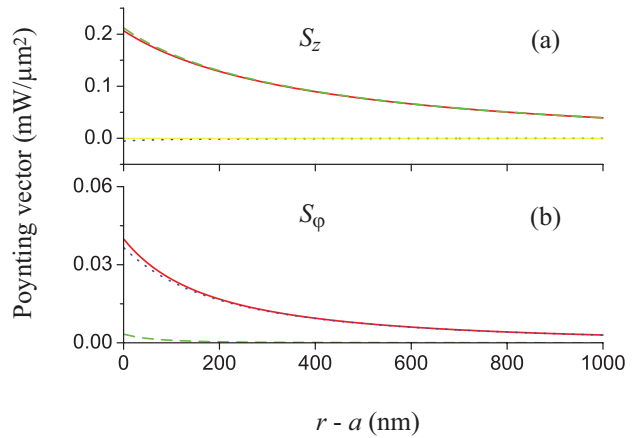}
 \end{center}
\caption{(Color online) Same as Fig.~\ref{fig2} but the surrounding medium is water, with the refractive index $n_2=1.33$.
}
\label{fig3}
\end{figure}

An important characteristic of the light propagation is the cycle-averaged Poynting vector
\begin{equation}
\mathbf{S}=\frac{1}{2} 
\mathrm{Re}[\boldsymbol{\mathcal{E}}\times\boldsymbol{\mathcal{H}}^*].
\label{p13}
\end{equation}
We denote the axial, azimuthal, and radial components of the vector $\mathbf{S}$ in the cylindrical coordinates by the notation $S_z$, $S_{\varphi}$, and $S_r$, respectively. 
For quasicircularly polarized guided modes of fibers, we have $S_r=0$. 
Outside the fiber, that is, for $r>a$, the axial and azimuthal components are given by \cite{angular}
\begin{eqnarray}
S_z&=&|\mathcal{N}|^2\frac{\omega\epsilon_0 n_2^2}{\beta}
[(1-s)(1-s_2)K_0^2(qr)\nonumber\\
&&\mbox{}+(1+s)(1+s_2)K_2^2(qr)],\nonumber\\
S_{\varphi}&=&|\mathcal{N}|^2\frac{\omega\epsilon_0 n_2^2q}{\beta^2}
[(1-2s_2+s_2s)K_0(qr)\nonumber\\
&&\mbox{}-(1+2s_2+s_2s)K_2(qr)]K_1(qr),
\label{p18}
\end{eqnarray}
where the parameters $q$, $s$, and $s_2$ are given in Appendix \ref{sec:mode}.
The axial component $S_z$ 
describes the energy flow that propagates along the fiber.
The azimuthal component $S_{\varphi}$ 
describes the energy flow that circulates around the fiber. The presence of this flow is due to the existence of the longitudinal components $\mathcal{E}_z$ and 
$\mathcal{H}_z$ of the field in the fundamental mode. The trajectory of the Poynting vector
is described by the spiral curve $(r,\varphi,z)$ 
where $r=r_0$ and $d\varphi/dz=S_{\varphi}/rS_z$, with $r_0$ being a constant \cite{AllenOptCommun}. Since $S_z$
and $S_{\varphi}$ do not depend on $z$, the rotation angle of the Poynting vector
is given by $\varphi=\varphi(0)+k_S z$, where $k_S=S_{\varphi}/rS_z$.  
The period of the trajectory of the Poynting vector along the $z$ axis is $2\pi/k_S$.
The trajectory is illustrated in Fig.~\ref{fig1}. The radial dependencies of the components $S_z$ and $S_\varphi$
are plotted in Fig.~\ref{fig2} for the case of a nanofiber in vacuum and in Fig.~\ref{fig3} for the case of a nanofiber in water. The solid red lines of these figures show that $S_z$ and $S_\varphi$ are positive, that is, the energy of the counterclockwise polarized guided light field propagates in the forward direction and circulates counterclockwise around the nanofiber. Our additional numerical calculations which are not shown here confirm the result of Ref.~\cite{Mokhov06} that the axial component $S_z$ may become negative when the refractive index $n_1$ of the fiber is large enough ($n_1/n_2>2.71$).
However, we are not interested in such high-index fibers. We note that the existence of the azimuthal component $S_\varphi$ in the case of guided fields is similar to that in the case of light beams with a transverse phase gradient \cite{Roichman08}. Such a component leads to a force transverse to the direction of propagation. The observation of
optical forces arising from transverse phase gradients has been experimentally demonstrated \cite{Roichman08}.

The propagation power $P_z$ is determined as the integral of $S_z$ over the transverse plane of the fiber, that is, 
\begin{equation}
P_z=\int S_z\,d^2\mathbf{r}.
\label{p14}
\end{equation}
Here we have introduced the notation $\int d^2\mathbf{r}=\int_0^{2\pi}d\varphi\int_0^{\infty}r\,dr$. 
When the fiber material is nonabsorbing and nondispersive,
the energy per unit length is given by 
\begin{equation}
U=\frac{\epsilon_0}{2}\int n^2|\boldsymbol{\mathcal{E}}|^2 \,d^2\mathbf{r}.
\label{p15}
\end{equation}
Here $n(r)=n_1$ and $n_2$ for $r<a$ and $r>a$, respectively.
We note that the propagation power $P_z$ is related to  the energy per unit length $U$
by the formula 
\begin{equation}
P_z=Uv_g,
\end{equation} 
where $v_g=1/\beta'(\omega)\equiv(d\beta/d\omega)^{-1}$ is the group velocity of light in the guided mode.

According to \cite{Bliokh13},
the momentum density of the field can be decomposed into two parts, the orbital part and the spin part.
We apply such a decomposition to the Pointing vector. Then, we obtain
\begin{equation}\label{p38}
\mathbf{S}=\mathbf{S}^{\mathrm{orb}}+\mathbf{S}^{\mathrm{spin}},
\end{equation}
where 
\begin{equation}\label{p39}
\mathbf{S}^{\mathrm{orb}}=\frac{c\epsilon_0}{2k}\mathrm{Im}[\boldsymbol{\mathcal{E}}^*\cdot(\boldsymbol{\nabla})\boldsymbol{\mathcal{E}}] 
\end{equation}
and
\begin{equation}\label{p40}
\mathbf{S}^{\mathrm{spin}}=\frac{c\epsilon_0}{4k}\boldsymbol{\nabla}\times\mathrm{Im}[\boldsymbol{\mathcal{E}}^*\times\boldsymbol{\mathcal{E}}]
\end{equation}
are the orbital and spin parts, respectively. Here, 
$k=2\pi/\lambda$ is the wave number of the light field in free space.
In Eq.~(\ref{p39}), the dot product applies to the vectors $\boldsymbol{\mathcal{E}}^*$ and $\boldsymbol{\mathcal{E}}$, that is, we use the notation
$\boldsymbol{\mathcal{E}}^*\cdot(\boldsymbol{\nabla})\boldsymbol{\mathcal{E}}\equiv \sum_{i=x,y,z}\mathcal{E}_i^*\boldsymbol{\nabla}\mathcal{E}_i$. 
For the forward counterclockwise quasicircularly polarized guided field given by Eq.~(\ref{p9}), we find
\begin{eqnarray}\label{p41}
S_z^{\mathrm{orb}}&=&\frac{c\epsilon_0\beta}{2k}|\boldsymbol{\mathcal{E}}|^2,\nonumber\\
S_\varphi^{\mathrm{orb}}&=&
\frac{c\epsilon_0}{2kr}[|\mathcal{E}_z|^2+|\mathcal{E}_r+i\mathcal{E}_\varphi|^2]
=\frac{c\epsilon_0}{2kr}[|\mathcal{E}_z|^2+2|\mathcal{E}_1|^2],
\nonumber\\
\end{eqnarray}
and
\begin{eqnarray}\label{p42}
S_z^{\mathrm{spin}}&=&\frac{c\epsilon_0}{2kr}\frac{\partial}{\partial r}
[r\mathrm{Im}(\mathcal{E}_r\mathcal{E}_z^*)],\nonumber\\
S_\varphi^{\mathrm{spin}}&=&-\frac{c\epsilon_0}{2k}\frac{\partial}{\partial r} \mathrm{Im}(\mathcal{E}_\varphi\mathcal{E}_r^*).
\end{eqnarray}
Here we have used the notation
$\mathcal{E}_{\pm 1}=\mp(\mathcal{E}_x \pm i\mathcal{E}_y)/\sqrt2=\mp(\mathcal{E}_r \pm i\mathcal{E}_\varphi)e^{\pm i\varphi}/\sqrt2$ for the spherical tensor components of the field. 

Equations (\ref{p41}) show that the orbital parts $S_z^{\mathrm{orb}}$ and $S_\varphi^{\mathrm{orb}}$ of the axial and azimuthal components of the Poynting vector, respectively, are always positive (see the dashed green curves in Figs.~\ref{fig2} and \ref{fig3}). We note that the orbital part $S_\varphi^{\mathrm{orb}}$ of the azimuthal component of the Poynting vector is produced by the longitudinal field component $\mathcal{E}_z$ and the spherical tensor component 
$\mathcal{E}_1$. The latter characterizes the clockwise polarization that is opposite to the principal counterclockwise polarization of the field in the case considered. 

Equations (\ref{p42}) show that the spin parts $S_z^{\mathrm{spin}}$ and $S_\varphi^{\mathrm{spin}}$ of the axial and azimuthal components of the Poynting vector, respectively, can be either positive or negative depending on the parameters.
We observe that, outside the fiber, for our parameters, 
the spin part $S_z^{\mathrm{spin}}$ of the axial component $S_z$
is negative [see the dotted blue curves in Figs.~\ref{fig2}(a) and \ref{fig3}(a)], while 
the spin part $S_\varphi^{\mathrm{spin}}$ of the azimuthal component $S_\varphi$
is positive [see the dotted blue curves in Figs.~\ref{fig2}(b) and \ref{fig3}(b)]. 
It is clear that the presence of the spin part $S_z^{\mathrm{spin}}$ of the axial component $S_z$ results from the presence of the longitudinal component $\mathcal{E}_z$ of the guided light field. 
When the refractive index contrast $n_1/n_2$ is large enough, the magnitude of the negative spin part $S_z^{\mathrm{spin}}$ outside the fiber may exceed that of the positive orbital part $S_z^{\mathrm{orb}}$. In this case, the axial component $S_z$ becomes negative \cite{Mokhov06}. A similar situation occurs in the case of Bessel beams \cite{Novitsky07}.

Comparison between Figs.~\ref{fig2} and \ref{fig3} shows that the magnitudes of the components of the Poynting vector 
in the presence of water are smaller and reduce with increasing the distance $r-a$ slower than in vacuum. 
The reason is that, when immersed in water, the size parameter $V=ka\sqrt{n_1^2-n_2^2}$ 
for a nanofiber with radius of 250 nm is small because
of the reduction of the contrast in refractive index between the core
and the surrounding medium. For such a nanofiber, the guided field can penetrate deeply into the outside, making its spatial gradient small. In addition, the longitudinal component of the field is small, the axial Poynting vector component $S_z$ consists of mainly the orbital part [see the solid red curve and the dashed green curve of Fig.~\ref{fig3}(a)], while
the azimuthal component $S_\varphi$ consists of mainly the spin part [see the solid red curve and the dotted blue curve of Fig.~\ref{fig3}(b)]. 

\section{Force of the nanofiber-guided light field on a dielectric particle}
\label{sec:force}

We consider the scattering of the guided light field from a dielectric spherical particle in the vicinity of the nanofiber in vacuum or in a dielectric medium.
Let $a_s$ and $\bar{n}=\sqrt{\epsilon}$ be the radius and refractive index of the particle, respectively, 
with $\epsilon$ being the dielectric constant of the particle material. Various formulations of 
the generalized Lorentz-Mie theory for a spherical particle in an arbitrary incident light field have been developed \cite{Barton88,Gouesbet88,Barton89,GLMT,Salandrino12}. The formulation of the generalized theory by Barton \textit{et al.} \cite{Barton88} is summarized in Appendix \ref{sec:Mie}. We use this formulation in our numerical calculations. We omit multiple scattering between the particle and the fiber surface. This approximation is valid when the particle is small or not too close to the fiber surface or when the contrast in refractive index between the fiber and the medium is not too high. 

In the steady-state condition, the radiation force $\mathbf{F}$ on the particle is given by the formula \cite{Jackson} 
\begin{equation}\label{p34}
\mathbf{F}=\langle\oint\limits_{S} \hat{\mathbf{n}}\cdot \overleftrightarrow{T} dS\rangle,
\end{equation}
which is the time average of the integral of the scalar product of the outwardly directed normal unit vector $\hat{\mathbf{n}}$ and the Maxwell's stress tensor $\overleftrightarrow{T}$ over a surface enclosing the particle.
The radiation force of an arbitrary incident light field on a spherical particle has been calculated \cite{Barton89,GLMT,Salandrino12}. According to \cite{Barton89},
the components of the radiation force $\mathbf{F}$ are given by
\begin{eqnarray}\label{p36}
F_x+iF_y&=&\frac{i\epsilon_0n_2^2k^2}{4}\sum_{l=1}^{\infty}\sum_{m=-l}^{l}\bigg\{l(l+2)
\nonumber\\&&\mbox{}
\times\sqrt{\frac{(l+m+2)(l+m+1)}{(2l+1)(2l+3)}}
\nonumber\\&&\mbox{}
\times(2n_2^2a_{lm}a_{l+1,m+1}^*+n_2^2a_{lm}A_{l+1,m+1}^*
\nonumber\\&&\mbox{}
+n_2^2A_{lm}a_{l+1,m+1}^*+2b_{lm}b_{l+1,m+1}^*
\nonumber\\&&\mbox{}
+b_{lm}B_{l+1,m+1}^*+B_{lm}b_{l+1,m+1}^*)
\nonumber\\&&\mbox{}
+l(l+2)\sqrt{\frac{(l-m+2)(l-m+1)}{(2l+1)(2l+3)}} 
\nonumber\\&&\mbox{}
\times(2n_2^2a_{l+1,m-1}a_{lm}^*+n_2^2a_{l+1,m-1}A_{lm}^*
\nonumber\\&&\mbox{}
+n_2^2A_{l+1,m-1}a_{lm}^*+2b_{l+1,m-1}b_{lm}^*
\nonumber\\&&\mbox{}
+b_{l+1,m-1}B_{lm}^*+B_{l+1,m-1}b_{lm}^*) 
\nonumber\\&&\mbox{}
+\sqrt{(l+m+1)(l-m)}  n_2(2a_{lm}b_{l,m+1}^*
\nonumber\\&&\mbox{}
+a_{lm}B_{l,m+1}^*+A_{lm}b_{l,m+1}^*-2b_{lm}a_{l,m+1}^*
\nonumber\\&&\mbox{}
-b_{lm}A_{l,m+1}^*-B_{lm}a_{l,m+1}^*)\bigg\}
\end{eqnarray}
and
\begin{eqnarray}\label{p37}
F_z&=&-\frac{\epsilon_0n_2^2k^2}{2}\sum_{l=1}^{\infty}\sum_{m=-l}^{l} \mathrm{Im}\bigg\{l(l+2) 
\nonumber\\&&\mbox{}
\times \sqrt{\frac{(l-m+1)(l+m+1)}{(2l+1)(2l+3)}}
(2n_2^2a_{l+1,m}a_{lm}^*
\nonumber\\&&\mbox{}
+n_2^2a_{l+1,m}A_{lm}^*+n_2^2A_{l+1,m}a_{lm}^*
+2b_{l+1,m}b_{lm}^*
\nonumber\\&&\mbox{}
+b_{l+1,m}B_{lm}^*+B_{l+1,m}b_{lm}^*)
\nonumber\\&&\mbox{}
+mn_2(2a_{lm}b_{lm}^*+a_{lm}B_{lm}^*+A_{lm}b_{lm}^*)\bigg\}.
\end{eqnarray}
Here, $A_{lm}$ and $B_{lm}$ are the beam shape coefficients, and $a_{lm}$ and $b_{lm}$ are the scattering coefficients \cite{Barton88,Barton89}.
The expressions for these coefficients are given in Appendix \ref{sec:Mie}, where the notation $a$ is used, instead of $a_s$, for the radius of the dielectric spherical particle,
and $r$, $\theta$, and $\varphi$ are the spherical coordinate system with the origin at the center of the particle.

For a Rayleigh particle having a radius $a_s$ very small as compared to the wavelength $\lambda$ of light ($a_s\ll\lambda$), the electric dipole approximation can be used \cite{Mie books}.
In the framework of this approximation, the particle can be considered as an electric dipolar particle. The electric polarizability of the particle is given by $\alpha=6\pi i\epsilon_0 a_1^{\mathrm{Mie}}/n_2k^3$, where $a_l^{\mathrm{Mie}}$ is the  conventional Mie coefficient given by Eq.~(\ref{p28b}). In the limit $ka_s\ll1$ and in the absence of absorption (for real $\bar{n}$),  we have \cite{Mie books}
\begin{eqnarray}\label{p48}
\mathrm{Re}(\alpha)&\simeq&  4\pi\epsilon_0 n_2^2 a_s^3 \frac{\bar{n}^2-n_2^2}{\bar{n}^2+2n_2^2},
\nonumber\\
\mathrm{Im}(\alpha)&\simeq&\frac{8\pi\epsilon_0}{3} n_2^5 k^3 a_s^6\bigg(\frac{\bar{n}^2-n_2^2}{\bar{n}^2+2n_2^2}\bigg)^2.
\end{eqnarray}
According to \cite{Salandrino12,Landau,Arias-Gonzalez03,Wong06,Albaladejo09,Nieto-Vesperinas10,Gomez-Medina12}, the force on the particle is
\begin{equation}\label{p43}
\mathbf{F}=\mathbf{F}^{\mathrm{grad}}+\mathbf{F}^{\mathrm{scat}},
\end{equation}
where
\begin{equation}\label{p44}
\mathbf{F}^{\mathrm{grad}}=\frac{1}{4}\mathrm{Re}(\alpha)\boldsymbol{\nabla} |\boldsymbol{\mathcal{E}}|^2
\end{equation}
is the gradient force and
\begin{equation}\label{p45}
\mathbf{F}^{\mathrm{scat}}=\sigma_{\mathrm{ext}}\frac{n_2}{c}\mathbf{S}-\sigma_{\mathrm{ext}}\frac{c}{2n_2}[\boldsymbol{\nabla}\times \mathbf{J}^\mathrm{spin}]
\end{equation}
is the scattering force. Here, the quantity
\begin{equation}\label{p45a} 
\sigma_{\mathrm{ext}}=\frac{k}{\epsilon_0n_2}\mathrm{Im}(\alpha)
\simeq \frac{8\pi}{3} n_2^4 k^4 a_s^6\bigg(\frac{\bar{n}^2-n_2^2}{\bar{n}^2+2n_2^2}\bigg)^2 
\end{equation}
is the extinction cross section of the Rayleigh particle \cite{Mie books}, and
\begin{equation}\label{p46}
\mathbf{J}^\mathrm{spin}=\frac{\epsilon_0 n_2^2}{2\omega} \mathrm{Im}[\boldsymbol{\mathcal{E}}^*\times\boldsymbol{\mathcal{E}}]
\end{equation}
is the spin density of the light field \cite{MandelWolf,Cohen-Tannoudji}. 
In Eq.~(\ref{p45}), the first term, 
$\sigma_{\mathrm{ext}}n_2\mathbf{S}/c\equiv \mathbf{F}^{\mathrm{press}}$, is the traditional radiation pressure, while the second term, 
$-\sigma_{\mathrm{ext}}c[\boldsymbol{\nabla}\times \mathbf{J}^\mathrm{spin}]/2n_2\equiv \mathbf{F}^{\mathrm{curl}}$, is called the force from the curl of the spin angular momentum \cite{Albaladejo09,Nieto-Vesperinas10,Gomez-Medina12}. 

When we use the decomposition (\ref{p38}) for the Poynting vector, we can decompose the traditional radiation pressure $\mathbf{F}^{\mathrm{press}}$ into two parts, the orbital part $\sigma_{\mathrm{ext}}n_2\mathbf{S}^{\mathrm{orb}}/c$ and the spin part $\sigma_{\mathrm{ext}}n_2\mathbf{S}^{\mathrm{spin}}/c$.
It is interesting to note that $[\boldsymbol{\nabla}\times \mathbf{J}^\mathrm{spin}]/2n_2^2=\mathbf{S}^{\mathrm{spin}}/c^2$.
Hence, we find $\mathbf{F}^{\mathrm{curl}}+\sigma_{\mathrm{ext}}n_2\mathbf{S}^{\mathrm{spin}}/c=0$. This formula means that
the force from the curl of the spin angular momentum cancels the spin part of the radiation pressure.
Consequently, Eq.~(\ref{p45}) can be rewritten in the form \cite{Bliokh13}
\begin{equation}\label{p47}
\mathbf{F}^{\mathrm{scat}}=\sigma_{\mathrm{ext}}\frac{n_2}{c}\mathbf{S}^{\mathrm{orb}}.
\end{equation}
Thus, the scattering force acting on a dipolar particle is proportional to the orbital part of the Poynting vector. The scattering force (\ref{p47}) can be interpreted as the orbital component of the radiation pressure force or simply as the radiation pressure in a more accurate form \cite{Bliokh13}. This interpretation is different from but equivalent to the traditional interpretation given in Refs.~\cite{Albaladejo09,Nieto-Vesperinas10,Gomez-Medina12}. We note that the definition used in Refs.~\cite{Albaladejo09,Nieto-Vesperinas10,Gomez-Medina12} for the spin density of the light field is not consistent with the textbook definition \cite{MandelWolf,Cohen-Tannoudji}. Indeed, a factor of $-2$ must be added to the former to remove the difference between the two definitions.

We use Eqs.~(\ref{p36}) and (\ref{p37}) to perform numerical calculations for the radiation force $\mathbf{F}$ on the particle.
We note that the analytical and numerical calculations from these equations are in perfect agreement with the calculations from Eq.~(\ref{p34}), which is the integral of the Maxwell's stress tensor $\overleftrightarrow{T}$ over a surface enclosing the particle. We present the results of our numerical calculations below.

We consider the situation where the particle is  positioned on the fiber surface. To be specific, we assume that the center of the particle is located on the axis $x$, that is, at the position with the coordinates $x=a+a_s$, $y=0$, and $z=0$ in the Cartesian coordinate frame of the fiber. With this specific choice, we have $F_x=F_r$ and $F_y=F_\varphi$. In our numerical calculations, we use a guided light field with the wavelength $\lambda=1064$ nm and the propagation power $P_z=1$ mW. We examine two different cases: the system is in vacuum and the system is immersed in water.

\subsection{In vacuum}

\begin{figure}[tbh]
\begin{center}
  \includegraphics{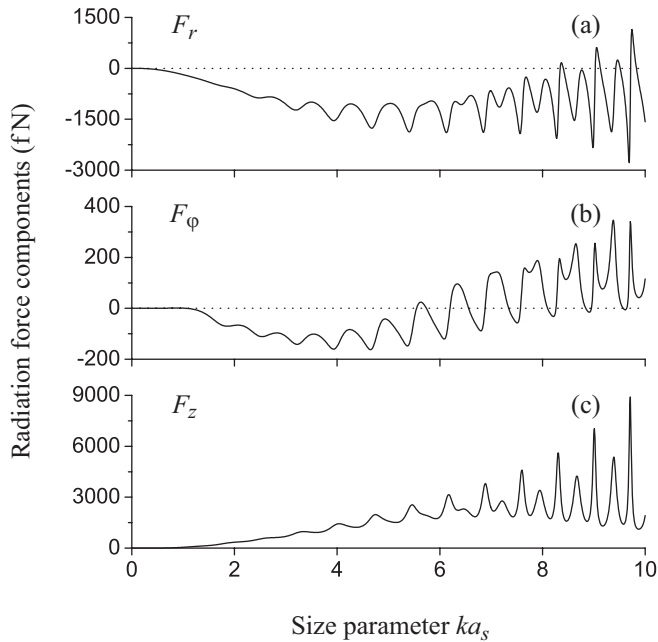}
 \end{center}
\caption{Components $F_r$ (a), $F_\varphi$ (b), and $F_z$ (c) of the force of the guided light as functions of the size parameter $ka_s$ of the particle in vacuum, whose refractive index is $n_2=1$.
The radius of the nanofiber is $a=250$ nm. The refractive index of the nanofiber is $n_1=1.45$. 
The wavelength of light is $\lambda=1064$ nm. The power of light is $P_z=1$ mW.
The particle-material refractive index is $\bar{n}=\sqrt{\epsilon}=\sqrt{2.6}$.
The particle is positioned on the fiber surface.
The dotted lines in parts (a) and (b) stand for the zero value of the force components and are guides to the eye. 
}
\label{fig4}
\end{figure}

\begin{figure}[tbh]
\begin{center}
  \includegraphics{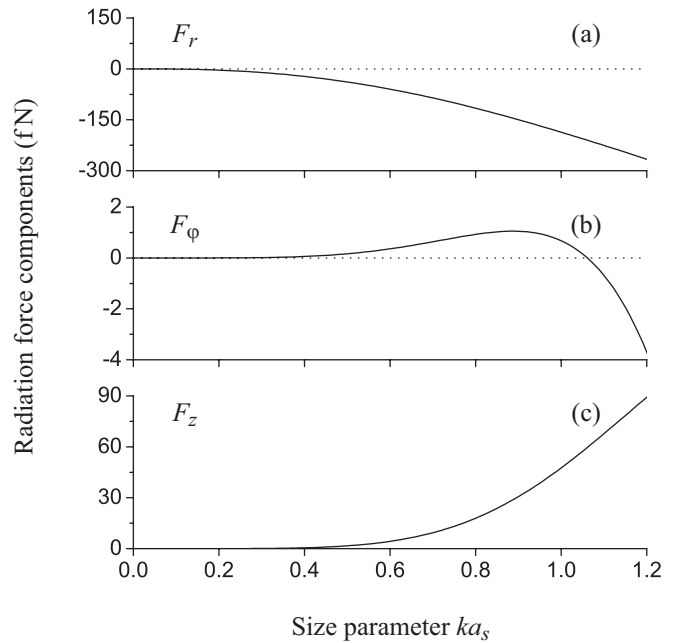}
 \end{center}
\caption{Details of Fig.~\ref{fig4} in the range $ka_s\leq 1.2$. 
}
\label{fig5}
\end{figure}

First, we consider the case where the medium around the fiber and the particle is vacuum, with the refractive index $n_2=1$.
In Fig.~\ref{fig4}, we plot the radial, azimuthal, and axial components $F_r$, $F_\varphi$, and $F_z$, respectively, of the radiation force as functions of the size parameter $ka_s$ of the particle in the range $ka_s\leq 10$. In these calculations, we use the fiber radius $a=250$ nm, the fiber refractive index $n_1=1.45$ (for silica), and 
the particle-material refractive index $\bar{n}=\sqrt{\epsilon}=\sqrt{2.6}$ 
(for a polystyrene particle). The details of Fig.~\ref{fig4} in the range $ka_s\leq 1.2$ are shown in Fig.~\ref{fig5}.

We observe from Fig. \ref{fig4} the oscillations in the dependencies of the force components on the size parameter $ka_s$ of the particle. The oscillatory behavior is evident when $ka_s$ is not small. Such oscillations occur due to the Mie resonances associated with the whispering gallery modes of the spherical particle \cite{Mie books}. We observe from Fig.~\ref{fig4}(a) that $F_r<0$ in the region $ka_s\leq 8$ and $F_r>0$ in some narrow intervals in the region $ka_s>8$.
The negative or positive values of the radial force component $F_r$ indicate that the particle is attracted to or repelled from the fiber surface, respectively, depending on the size of the particle. The repulsion in some narrow intervals of large $ka_s$ is due to the excitation of and interference between circulating whispering-gallery modes that may become totally reflected from the boundary at the upper (remotest) part of the sphere \cite{Almaas13}. We observe from Fig.~\ref{fig4}(c) that $F_z>0$ in the whole range ($ka_s\leq 10$) of the figure. The positive axial force component $F_z$ means that the particle is pushed along the propagation direction. 
Similar features have been observed in the case of a particle in the evanescent wave produced by a flat surface \cite{Almaas95,Almaas13}. 

Figure \ref{fig4}(b) shows that the azimuthal force component $F_\varphi$  becomes negative in a wide region of $ka_s$, namely, in the region $1.06< ka_s< 5.55$. Such a negative force is directed oppositely to the direction of the azimuthal Poynting vector component $S_\varphi$, that is, oppositely to the direction of the energy circulation around the nanofiber [see the solid red curve in Fig.~\ref{fig2}(b)]. The negative or positive values of the azimuthal force component $F_\varphi$ mean that the particle undergoes a negative or positive torque, respectively, depending on the size of the particle. Such a negative azimuthal force occurs in the case where the incident quasicircularly polarized guided field is scattered dominantly in the positive direction of the azimuthal energy flow and, based on the principle of action and reaction, the momentum transfer leads to an azimuthally backward force, similar to the case of axial drag forces \cite{Chen11,Novitsky11}. The dominant scattering in the direction of the energy-flow circulation is a result of interference between multipolar fields in the particle. When the radius $a_s$ of the particle is very small as compared to the wavelength $\lambda$, the particle is dipolar and, consequently, the difference
between forward and backward scattering is negligible. Then, the force is mainly determined by the momentum removed from the incident field. In the case considered here, where $S_\varphi$ is positive at the position of the particle, the removed momentum is positive in the azimuthal direction. This is the reason why Fig.~\ref{fig5}(b) shows $F_\varphi>0$ for $ka_s<1.06$. 
 
\begin{figure}[tbh]
\begin{center}
  \includegraphics{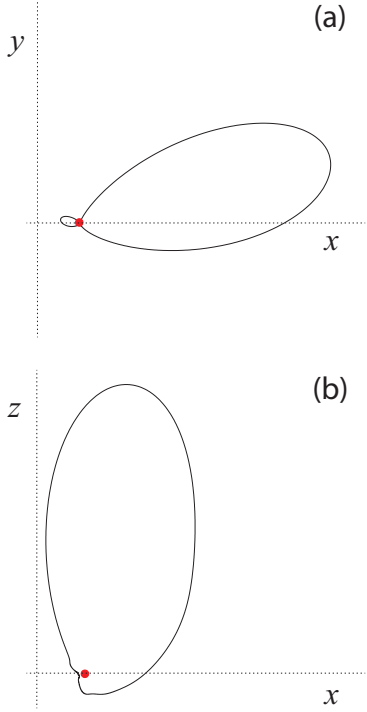}
 \end{center}
\caption{(Color online) Diagrams for scattering in the fiber transverse plane $(x,y)$ and the axial plane $(x,z)$.
The particle (red circle) is positioned on the fiber surface at the $x$ axis. The size parameter of the particle is $ka_s=4$.
Other parameters are as in Fig.~\ref{fig4}. 
}
\label{fig6}
\end{figure}

The diagrams for scattering in the fiber transverse plane $(x,y)$ and the axial plane $(x,z)$ are shown in Fig.~\ref{fig6}. We observe from Fig.~\ref{fig6} that, for $ka_s=4$, the scattering is dominant in the positive directions $+x$, $+y$, and $+z$. We note that, although the scattering is dominant in the $+z$ direction, the axial force component $F_z$ is positive. 
The reason is that, due to the significant axial component $S_z$ of the Poynting vector, the radiation momentum imparted to the particle in the $+z$ direction is large.

\subsection{In water}

\begin{figure}[tbh]
\begin{center}
  \includegraphics{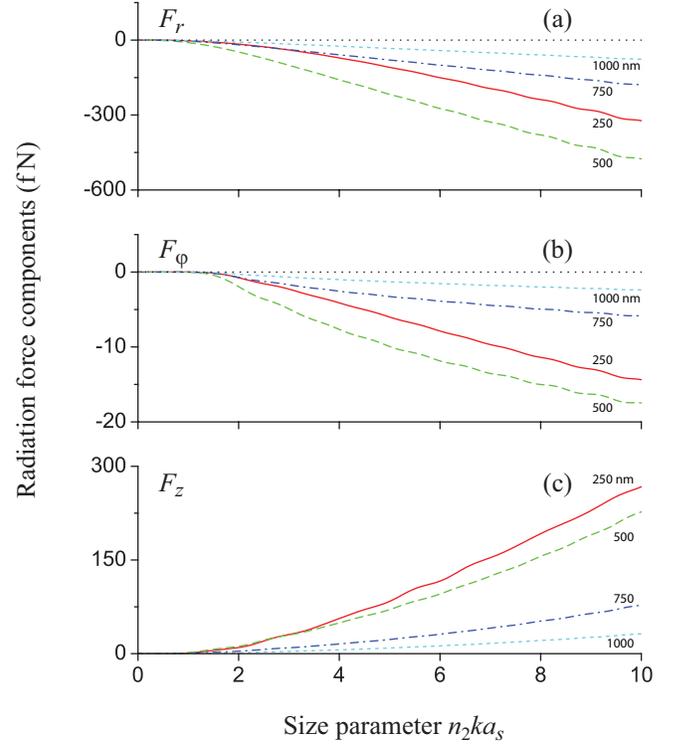}
 \end{center}
\caption{(Color online)
Components $F_r$ (a), $F_\varphi$ (b), and $F_z$ (c) of the force of the guided light as functions of the size parameter $n_2ka_s$ of the particle immersed in water, whose refractive index is $n_2=1.33$. The radius of the nanofiber is $a=250$ nm (solid red lines), 500 nm (dashed green lines), 750 nm (dash-dotted blue lines), and 1000 nm (dotted cyan lines). 
Other parameters are the same as those for Fig.~\ref{fig4}:
$\lambda=1064$ nm, $P_z=1$ mW, $n_1=1.45$, and $\bar{n}=\sqrt{\epsilon}=\sqrt{2.6}$. 
The particle is positioned on the fiber surface. The thin dotted black lines in parts (a) and (b) stand for the zero value of the force components and are guides to the eye. 
}
\label{fig7}
\end{figure}

\begin{figure}[tbh]
\begin{center}
  \includegraphics{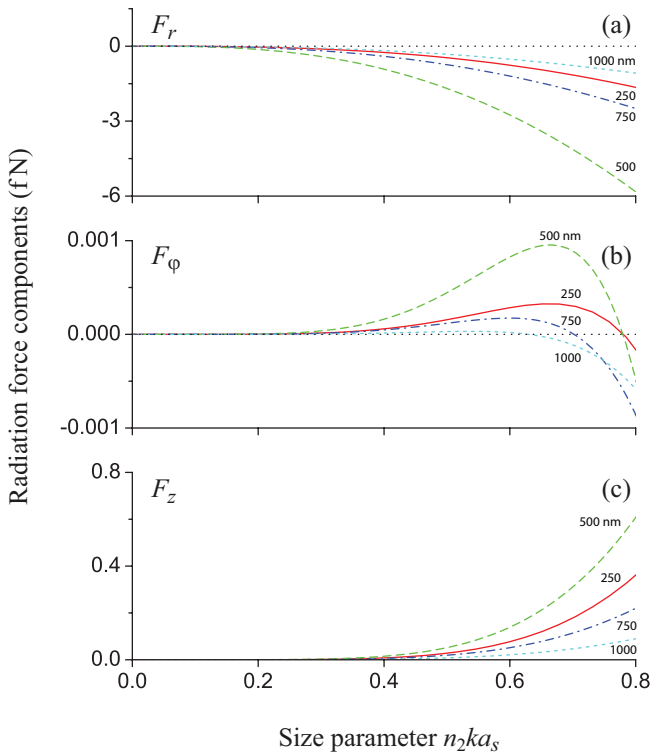}
 \end{center}
\caption{(Color online)
Details of Fig.~\ref{fig7} in the range $n_2ka_s\leq 0.8$. 
}
\label{fig8}
\end{figure}

We now consider the case where the fiber and the particle are immersed in water, whose refractive index is $n_2=1.33$.
In Fig.~\ref{fig7}, we plot the components of the radiation force as functions of the size parameter $n_2ka_s$ of the particle in water  in the range $ka_s\leq 10$.  The radius of the nanofiber is 250 nm (red lines), 500 nm (green lines), 750 nm (blue lines), and 1000 nm (cyan lines). Other parameters are as for Fig.~\ref{fig4}.
The details of Fig.~\ref{fig7} in the range $ka_s\leq 0.8$ are shown in Fig.~\ref{fig8}.

Figure \ref{fig7} shows that the oscillatory behavior in the dependencies of the force components on the size parameter of the particle becomes very weak when the system is immersed in water. 
In addition, the magnitude of the force in the case of water is smaller than that in the case of vacuum. 
These features are mainly due to the reduction of the contrast in refractive index between the particle and the surrounding medium. The reduction of the size parameter $V=ka\sqrt{n_1^2-n_2^2}$ of the fiber and the reduction of the contrast in refractive index between the nanofiber (the core) and the surrounding medium (the cladding)
also play a role. Indeed, the presence of water makes the fiber size parameter $V$ smaller, the field confinement weaker, and the spread of the evanescent wave field in the outside of the fiber broader. These features partially contribute to the reduction of the force of the guided light on the particle. 

It is worth noting that, according to the green line of Fig.~\ref{fig7}(b), for a polystyrene particle with radius of 410~nm 
($n_2ka_s\simeq 3.2$), a nanofiber with radius of 500 nm, and a guided light field with power of 1 mW, we obtain
a negative azimuthal force $F_\varphi\simeq-5.6$ fN. By increasing the power of light to 60 mW, we can achieve the force value 
$F_\varphi\simeq -0.34$ pN. This value is almost the same as that of the negative axial force realized in the interfering plane-wave experiment \cite{Brzobohaty13}.

We observe from Figs.~\ref{fig7}(a) and \ref{fig7}(c) 
that $F_r<0$ and $F_z>0$, respectively, in the whole range $n_2ka_s\leq 10$ of the figures.
Figures~\ref{fig7}(b) and \ref{fig8}(b) show that the azimuthal force component is negative, i.e. $F_\varphi<0$, in a broad range $0.8<n_2ka_s<10$ of the
size parameter of the particle. This is a result of the reduction of the contrast in refractive index between the particle and the surrounding medium. Indeed, a decrease in the ratio $\bar{n}/n_2$ between the refractive index $\bar{n}$ of the particle and the refractive index $n_2$ of the surrounding medium leads to an increase  
in the widths of the intervals of appropriate values of $n_2ka_s$ in which the particle can scatter the incident light dominantly in the positive (forward) azimuthal direction of the energy circulation. The manifestation of such a broadening is related to the reduction of the $Q$ factor of the Mie resonances. 
Another favorable factor for the azimuthal force $F_\varphi$ to be negative in a broad interval of $n_2ka_s$ 
is that the orbital azimuthal Poynting vector part $S_\varphi^{\mathrm{orb}}$, which determines the radiation pressure on a dipolar particle in the $\varphi$ direction, is very small in a broad range of the distance $r-a$ [see the dashed green line in Fig.~\ref{fig3}(b)].   

Comparison between the red, green, blue, and cyan lines of Figs.~\ref{fig7} and \ref{fig8}
shows that variations in the fiber radius affect the components of the force in a complicated way. 
Indeed, we observe that, when the fiber radius increases, the force components may either reduce or increase, depending on the region of parameters.

\begin{figure}[tbh]
\begin{center}
  \includegraphics{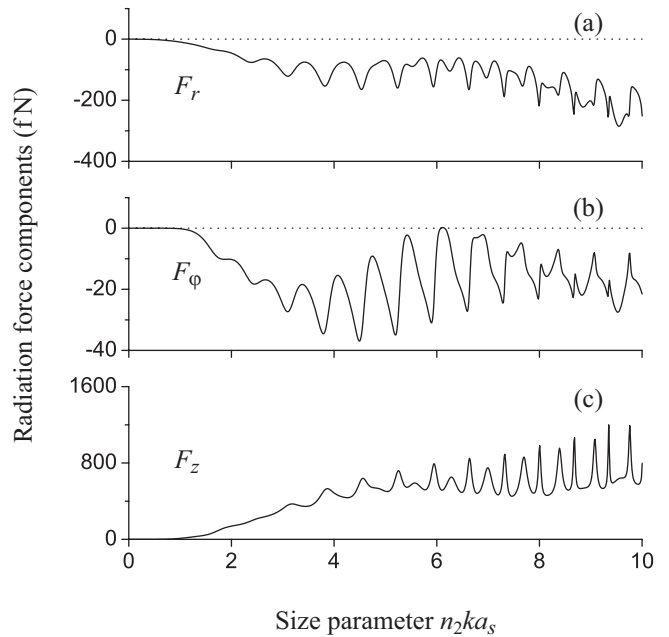}
 \end{center}
\caption{
Components $F_r$ (a), $F_\varphi$ (b), and $F_z$ (c) of the force of the guided light as functions of the size parameter $n_2ka_s$ of the particle in the case where 
the particle-material refractive index is $\bar{n}=\sqrt{\epsilon}=\sqrt{5}$ and the radius of the nanofiber is $a=250$ nm. Other parameters are the same as those for Fig.~\ref{fig7}.
}
\label{fig9}
\end{figure}

\begin{figure}[tbh]
\begin{center}
  \includegraphics{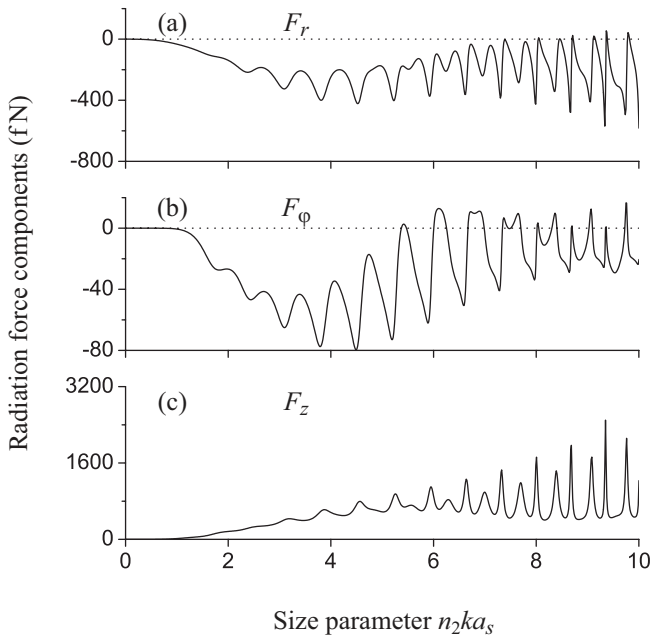}
 \end{center}
\caption{
Same as Fig.~\ref{fig9} except for the fiber radius $a=500$ nm.
}
\label{fig10}
\end{figure}

In order to see the effects of the refractive index $\bar{n}=\sqrt{\epsilon}$ of the material of the particle on the force,
we perform numerical calculations for the case where $\bar{n}=\sqrt{\epsilon}=\sqrt{5}$. In Figs.~\ref{fig9} and \ref{fig10}, we plot the components of the force from the guided light field of a nanofiber with the fiber radius of 250 nm and 500 nm, respectively.
We observe from both figures the oscillatory behavior that is typical in the scattering from Mie particles \cite{Mie books}.
Figure \ref{fig9} shows that, in the case of the fiber radius of 250 nm, the azimuthal force $F_\varphi$ is negative in a broad interval of the particle size parameter $n_2ka_s$. Meanwhile, in the case of Fig. \ref{fig10}, where the fiber radius is 500 nm, the azimuthal force $F_\varphi$ is negative in a smaller (but still significant) interval of $n_2ka_s$. Outside this interval, the sign of $F_\varphi$ sequentially changes from  negative to positive and vice versa. Comparison between Figs.~\ref{fig9} and \ref{fig10} shows that the oscillations and magnitudes of the dependencies of the force components on the particle size parameter $n_2ka_s$ do not depend much on the fiber radius $a$.
This stems from the fact that the Mie resonances are not modified significantly by the change of the fiber radius.

We recognize that multiple backscattering between the particle and the fiber surface on the force was omitted in our treatment. We note that the effect of multiple reflections between a particle and a dielectric surface has been studied in the framework of a ray-optic model for evanescent-wave scattering \cite{Prieve93}. Rigorous multiple scattering numerical methods have been developed \cite{Lester99,Chaumet00}. In the context of the experimental demonstration of a tractor beam, the influence of the scattered field reflected from a mirror back towards the particle has also been investigated \cite{Brzobohaty13}. According to the above mentioned studies, the effect of multiple scattering is not serious even for a particle located on the surface if the values of the size parameters are moderate. Indeed, it has been shown that, for a polystyrene
particle with a diameter of 7 $\mu$m in alcohol, located at a very small separation
distance (about 1 nm) from a MgF$_2$ film, more than 98\% of the scattered
energy is due to the initial contact of the evanescent wave \cite{Prieve93}. 
For smaller particles, such as those considered in
the present paper, the effect of multiple scattering on the force is expected to be very small.
The numerical calculations of Ref.~\cite{Brzobohaty13} for the experimental realization of a tractor beam have also confirmed that the reflection of the scattered field by a mirror does not significantly alter the resulting optical force. 
Therefore, we believe that the effect of multiple scattering is not important in our
present problem, where moderate values of the size parameters were used in calculations. 

In order to observe the occurrence of a negative azimuthal force experimentally, it is desirable to minimize the motion of the particle along the fiber axis. 
This can be done by using a standing-wave scheme instead of the running-wave configuration studied in this paper.
We have calculated the force in the case of a pair of counterpropagating guided fields with the same quasicircular polarization. 
We have found for this scheme that
a negative azimuthal force can also occur while the particle experiences an axial force toward the positions of either quasi-antinodes or quasi-nodes of the standing-wave structure depending on the particle size parameter.

\section{Summary}
\label{sec:summary}

We have studied the Poynting vector of a quasicircularly polarized guided light field of a nanofiber, and
have calculated the force of the field on a dielectric spherical particle outside the nanofiber.
We have shown that the orbital parts of the axial and azimuthal components of the Poynting vector are always positive while the spin parts can be either positive or negative. 
The presence of the spin part of the axial Poynting vector component is related to the presence of the longitudinal component of the guided light field. It is the source of the negative axial Poynting vector obtained in Ref.~\cite{Mokhov06} for high-contrast optical nanofibers. We have found that, when the size parameter of the particle is appropriate,
the azimuthal component of the force is directed oppositely to the circulation direction of the energy flow around the nanofiber. 
The occurrence of such a negative azimuthal force indicates the occurrence of a negative torque upon the particle, depending on the size of the particle. A negative azimuthal force occurs when the incident quasicircularly polarized guided field is scattered dominantly in the positive direction of the azimuthal energy flow. Unlike the case of the azimuthal direction, a negative force
along the axial direction could not be obtained in the range of the parameters we considered. The reason is that the incident photons have a significant orbital momentum aligned along the axial direction. 
Our results open the way to future research on sorting, manipulating, and controlling dielectric particles using nanofibers.

\begin{acknowledgments}
We thank P. Schneeweiss for helpful comments and discussions.
Financial support by the Austrian Science Fund (FWF; Lise Meitner project No. M 1501-N27 and SFB NextLite project No. F 4908-N23) is gratefully acknowledged.
\end{acknowledgments}

\appendix

\section{Fiber guided mode functions}
\label{sec:mode}

Consider a nanofiber that is a silica cylinder of radius $a$ and refractive index $n_1$ and is surrounded by an infinite background medium of refractive index $n_2$,
where $n_2<n_1$. The radius of the nanofiber is well below a given wavelength $\lambda$ of light. Therefore, the nanofiber supports only the hybrid fundamental modes HE$_{11}$ corresponding to the given wavelength \cite{fiber books}. The light field in such a mode is strongly guided. It penetrates into the outside of the nanofiber in the form of an evanescent wave carrying a significant fraction of energy \cite{fibermode}.
For a fundamental guided mode HE$_{11}$ of a light field of frequency $\omega$ (free-space wavelength $\lambda=2\pi c/\omega$ and free-space wave number $k=\omega/c$), the propagation constant $\beta$ is determined by the
fiber eigenvalue equation \cite{fiber books}
\begin{eqnarray}
\frac{J_0(h a)}{h a J_1(h a)}&=&
-\frac{n_1^2+n_2^2}{2n_1^2}\frac{K_1'(q a)}{q a K_1(q a)}+ \frac{1}{h^2 a^2}
\nonumber\\&&\mbox{}
-\Bigg[\left(\frac{n_1^2-n_2^2}{2n_1^2}\frac{K_1'(q a)}{q a K_1(q a)}\right)^2
\nonumber\\&&\mbox{}
+\frac{\beta^2}{n_1^2 k^2}\left(\frac{1}{q^2a^2}+\frac{1}{h^2a^2}\right)^2 \Bigg]^{1/2}.
\label{p1}
\end{eqnarray}
Here the parameters
\begin{equation}\label{p2} 
h=(n_1^2k^2-\beta^2)^{1/2}
\end{equation} 
and 
\begin{equation}\label{p3} 
q=(\beta^2-n_2^2k^2)^{1/2} 
\end{equation} 
characterize the fields inside and outside the fiber, respectively. The notations $J_n$ and $K_n$ stand for the Bessel functions of the first kind and the modified Bessel functions of the second kind, respectively. 

The cylindrical vector components of the mode-profile functions $\mathbf{e}(\mathbf{r})$ and $\mathbf{h}(\mathbf{r})$ 
of the electric and magnetic parts, respectively, of the fundamental guided mode \cite{fiber books} are given, 
for $r<a$, by
\begin{eqnarray}
e_{r}&=&i\frac{q}{h}\frac{K_1(qa)}{J_1(ha)}[(1-s)J_0(hr)-(1+s)J_2(hr) ],
\nonumber\\
e_{\varphi}&=&-\frac{q}{h}\frac{K_1(qa)}{J_1(ha)}[(1-s)J_0(hr)+(1+s)J_2(hr) ],
\nonumber\\
e_{z}&=& \frac{2q}{\beta}\frac{K_1(qa)}{J_1(ha)}J_1(hr),
\label{p4}
\end{eqnarray}
and
\begin{eqnarray}
h_r&=&\frac{\omega\epsilon_0 n_1^2q}{\beta h}\frac{K_1(qa)}{J_1(ha)}
[(1-s_1)J_0(hr)+(1+s_1)J_2(hr)],
\nonumber\\
h_\varphi&=&i\frac{\omega\epsilon_0 n_1^2q}{\beta h}\frac{K_1(qa)}{J_1(ha)}
[(1-s_1)J_0(hr)-(1+s_1)J_2(hr)],
\nonumber\\
h_z&=& i \frac{2q}{\omega\mu_0} s \frac{K_1(qa)}{J_1(ha)} J_1(hr),
\label{p5}
\end{eqnarray} 
and, for $r>a$, by
\begin{eqnarray}
e_{r}&=&i[(1-s)K_0(qr)+(1+s)K_2(qr) ],
\nonumber\\
e_{\varphi}&=&-[(1-s)K_0(qr)-(1+s)K_2(qr) ],
\nonumber\\
e_{z}&=& \frac{2q}{\beta}K_1(qr)
\label{p6}
\end{eqnarray}
and
\begin{eqnarray}
h_r&=&\frac{\omega\epsilon_0 n_2^2}{\beta}
[(1-s_2)K_0(qr)-(1+s_2)K_2(qr)],
\nonumber\\
h_\varphi&=&i\frac{\omega\epsilon_0 n_2^2}{\beta}
[(1-s_2)K_0(qr)+(1+s_2)K_2(qr)],
\nonumber\\
h_z&=&i\frac{2q}{\omega\mu_0}s K_1(qr).
\label{p7}
\end{eqnarray}
Here the parameter $s$ is defined as
\begin{equation}\label{p8} 
s=\frac{{1}/{h^2a^2}+{1}/{q^2a^2}}{{J_1^\prime (ha)}/{haJ_1(ha)}+{K_1^\prime (qa)}/{qaK_1(qa)}}.
\end{equation}
The parameters $s_1$ and $s_2$ are related to $s$ via the formulae 
$s_1=(\beta^2/k^2n_1^2)s$ and $s_2=(\beta^2/k^2n_2^2)s$.
We note that the axial components $e_{z}$ and $h_{z}$ are significant in the case of nanofibers \cite{fibermode}. This makes guided modes of nanofibers very different from plane-wave modes of free space and from guided modes of conventional (weakly guiding) fibers \cite{fibermode,fiber books}.

\section{Scattering of an arbitrary incident light field from a dielectric spherical particle}
\label{sec:Mie}

Consider the scattering of an arbitrary light beam from a dielectric spherical particle in a dielectric medium.
The field distribution is described by the generalized Lorentz-Mie theory \cite{Barton88,Gouesbet88}.
We review the formulation of the generalized theory by Barton \textit{et al.} \cite{Barton88}.

Let $a$ and $\bar{n}=\sqrt{\epsilon}$ be the radius and refractive index of the particle, respectively, 
with $\epsilon$ being the dielectric constant of the particle material, and let $n_2$ be the refractive index of the surrounding medium. The parameters $\epsilon$ and, consequently, $\bar{n}$ for the particle can, in general, take complex values, that is, the particle can be absorbing.
However, we assume that $n_2$ is real, that is, the surrounding medium is nonabsorbing. 
We use the spherical coordinate system $(r,\theta,\varphi)$ with the origin at the center of the particle. 

\subsection{Incident field}

According to \cite{Barton88,Barton89}, the complex amplitudes $\boldsymbol{\mathcal{E}}^{(i)}$ and $\boldsymbol{\mathcal{H}}^{(i)}$ of the positive-frequency parts of 
the electric and magnetic components, respectively, of an arbitrary incident field can, in the spherical coordinates, be decomposed as
\begin{eqnarray}\label{p23}
\mathcal{E}_r^{(i)}&=&\frac{1}{r^2}\sum_{l=1}^{\infty}\sum_{m=-l}^{l}l(l+1)A_{lm}S_l(n_2kr)Y_{lm}(\theta,\varphi),
\nonumber\\
\mathcal{E}_\theta^{(i)}&=&\frac{n_2k}{r}\sum_{l=1}^{\infty}\sum_{m=-l}^{l}\bigg(A_{lm}S'_l(n_2kr)\frac{\partial Y_{lm}(\theta,\varphi)}{\partial\theta}
\nonumber\\&&\mbox{}
-\frac{m}{n_2}B_{lm}S_l(n_2kr)\frac{Y_{lm}(\theta,\varphi)}{\sin\theta}\bigg),
\nonumber\\
\mathcal{E}_\varphi^{(i)}&=&\frac{n_2k}{r}\sum_{l=1}^{\infty}\sum_{m=-l}^{l}\bigg(im A_{lm}S'_l(n_2kr)\frac{Y_{lm}(\theta,\varphi)}{\sin\theta}
\nonumber\\&&\mbox{}
-\frac{i}{n_2}B_{lm}S_l(n_2kr)\frac{\partial Y_{lm}(\theta,\varphi)}{\partial\theta}\bigg),
\end{eqnarray}
and
\begin{eqnarray}\label{p24}
\mathcal{H}_r^{(i)}&=&\frac{c\epsilon_0}{r^2}\sum_{l=1}^{\infty}\sum_{m=-l}^{l}l(l+1)B_{lm}S_l(n_2kr)Y_{lm}(\theta,\varphi),
\nonumber\\
\mathcal{H}_\theta^{(i)}&=&\frac{c\epsilon_0 n_2k}{r}\sum_{l=1}^{\infty}\sum_{m=-l}^{l}\bigg(B_{lm}S'_l(n_2kr)\frac{\partial Y_{lm}(\theta,\varphi)}{\partial\theta}
\nonumber\\&&\mbox{}
+mn_2A_{lm}S_l(n_2kr)\frac{Y_{lm}(\theta,\varphi)}{\sin\theta}\bigg),
\nonumber\\
\mathcal{H}_\varphi^{(i)}&=&\frac{c\epsilon_0 n_2k}{r}\sum_{l=1}^{\infty}\sum_{m=-l}^{l}\bigg(im B_{lm}S'_l(n_2kr)\frac{Y_{lm}(\theta,\varphi)}{\sin\theta}
\nonumber\\&&\mbox{}
+in_2A_{lm}S_l(n_2kr)\frac{\partial Y_{lm}(\theta,\varphi)}{\partial\theta}\bigg).
\end{eqnarray}
Here, $S_l(x)=xj_l(x)=\sqrt{\pi x/2}J_{l+1/2}(x)$ is the first-kind Riccati-Bessel function, and $Y_{lm}$ is the spherical harmonic function, while
the beam shape coefficients $A_{lm}$ and $B_{lm}$ are determined by the surface integrals of 
the radial components $\mathcal{E}_r^{(i)}(a,\theta,\varphi)$ and $\mathcal{H}_r^{(i)}(a,\theta,\varphi)$ of the electric and magnetic fields, respectively, via the formulae
\begin{eqnarray}\label{p25}
A_{lm}&=&\frac{a^2}{l(l+1)S_l(n_2ka)}
\nonumber\\&&\mbox{}
\times\int\limits_0^{2\pi}\int\limits_0^{\pi} \mathcal{E}_r^{(i)}(a,\theta,\varphi) Y_{lm}^*(\theta,\varphi)  \sin\theta  d\theta d\varphi,
\nonumber\\
B_{lm}&=&\frac{a^2}{c\epsilon_0 l(l+1)S_l(n_2ka)}
\nonumber\\&&\mbox{}
\times\int\limits_0^{2\pi}\int\limits_0^{\pi} \mathcal{H}_r^{(i)}(a,\theta,\varphi) Y_{lm}^*(\theta,\varphi)  \sin\theta  d\theta d\varphi.
\end{eqnarray}

\subsection{Scattered field}

For the scattered field, the complex amplitudes $\boldsymbol{\mathcal{E}}^{(s)}$ and $\boldsymbol{\mathcal{H}}^{(s)}$ of the positive-frequency parts of 
the electric and magnetic components, respectively, are given in the spherical coordinate system by \cite{Barton88,Barton89} 
\begin{eqnarray}\label{p26}
\mathcal{E}_r^{(s)}&=&\frac{1}{r^2}\sum_{l=1}^{\infty}\sum_{m=-l}^{l}l(l+1)a_{lm}\xi_l^{(1)}(n_2kr)Y_{lm}(\theta,\varphi),
\nonumber\\
\mathcal{E}_\theta^{(s)}&=&\frac{n_2k}{r}\sum_{l=1}^{\infty}\sum_{m=-l}^{l}\bigg(a_{lm}\xi_l^{(1)\prime}(n_2kr)\frac{\partial Y_{lm}(\theta,\varphi)}{\partial\theta}
\nonumber\\&&\mbox{}
-\frac{m}{n_2}b_{lm}\xi_l^{(1)}(n_2kr)\frac{Y_{lm}(\theta,\varphi)}{\sin\theta}\bigg),
\nonumber\\
\mathcal{E}_\varphi^{(s)}&=&\frac{n_2k}{r}\sum_{l=1}^{\infty}\sum_{m=-l}^{l}\bigg(im a_{lm}\xi_l^{(1)\prime}(n_2kr)\frac{Y_{lm}(\theta,\varphi)}{\sin\theta}
\nonumber\\&&\mbox{}
-\frac{i}{n_2}b_{lm}\xi_l^{(1)}(n_2kr)\frac{\partial Y_{lm}(\theta,\varphi)}{\partial\theta}\bigg),
\end{eqnarray}
and
\begin{eqnarray}\label{p27}
\mathcal{H}_r^{(s)}&=&\frac{c\epsilon_0}{r^2}\sum_{l=1}^{\infty}\sum_{m=-l}^{l}l(l+1)b_{lm}\xi_l^{(1)}(n_2kr)Y_{lm}(\theta,\varphi),
\nonumber\\
\mathcal{H}_\theta^{(s)}&=&\frac{c\epsilon_0 n_2k}{r}\sum_{l=1}^{\infty}\sum_{m=-l}^{l}\bigg(b_{lm}\xi_l^{(1)\prime}(n_2kr)\frac{\partial Y_{lm}(\theta,\varphi)}{\partial\theta}
\nonumber\\&&\mbox{}
+mn_2a_{lm}\xi_l^{(1)}(n_2kr)\frac{Y_{lm}(\theta,\varphi)}{\sin\theta}\bigg),
\nonumber\\
\mathcal{H}_\varphi^{(s)}&=&\frac{c\epsilon_0 n_2k}{r}\sum_{l=1}^{\infty}\sum_{m=-l}^{l}\bigg(im b_{lm}\xi_l^{(1)\prime}(n_2kr)\frac{Y_{lm}(\theta,\varphi)}{\sin\theta}
\nonumber\\&&\mbox{}
+in_2a_{lm}\xi_l^{(1)}(n_2kr)\frac{\partial Y_{lm}(\theta,\varphi)}{\partial\theta}\bigg).
\end{eqnarray}
Here, $\xi_l^{(1)}(x)=S_l(x)-iC_l(x)=xh_l^{(1)}(x)=\sqrt{\pi x/2}H_{l+1/2}^{(1)}(x)$ is the complex Riccati-Bessel function, with $C_l(x)=-xy_l(x)=-\sqrt{\pi x/2}Y_{l+1/2}(x)$ being the second-kind Riccati-Bessel function.
The scattering coefficients $a_{lm}$ and $b_{lm}$ are determined by the formulae
\begin{equation}\label{p28}
\begin{split}
a_{lm}=\frac{n_2S'_l(\bar{n}ka)S_l(n_2ka)-\bar{n}S_l(\bar{n}ka)S'_l(n_2ka)}{\bar{n}S_l(\bar{n}ka)\xi_l^{(1)\prime}(n_2ka)-n_2S'_l(\bar{n}ka)\xi_l^{(1)}(n_2ka)} A_{lm},
\\
b_{lm}=\frac{\bar{n}S'_l(\bar{n}ka)S_l(n_2ka)-n_2S_l(\bar{n}ka)S'_l(n_2ka)}{n_2S_l(\bar{n}ka)\xi_l^{(1)\prime}(n_2ka)-\bar{n}S'_l(\bar{n}ka)\xi_l^{(1)}(n_2ka)} B_{lm}.
\end{split}
\end{equation}
Note that Eqs.~(\ref{p28}) can be rewritten in the form
\begin{equation}\label{p28a}
\begin{split}
a_{lm}=-a_{l}^{\mathrm{Mie}} A_{lm},
\\
b_{lm}=-b_{l}^{\mathrm{Mie}} B_{lm},
\end{split}
\end{equation}
where
\begin{equation}\label{p28b}
\begin{split}
a_{l}^{\mathrm{Mie}}=\frac{\bar{n}S_l(\bar{n}ka)S'_l(n_2ka)-n_2S'_l(\bar{n}ka)S_l(n_2ka)}{\bar{n}S_l(\bar{n}ka)\xi_l^{(1)\prime}(n_2ka)-n_2S'_l(\bar{n}ka)\xi_l^{(1)}(n_2ka)},
\\
b_{l}^{\mathrm{Mie}}=\frac{n_2S_l(\bar{n}ka)S'_l(n_2ka)-\bar{n}S'_l(\bar{n}ka)S_l(n_2ka)}{n_2S_l(\bar{n}ka)\xi_l^{(1)\prime}(n_2ka)-\bar{n}S'_l(\bar{n}ka)\xi_l^{(1)}(n_2ka)}
\end{split}
\end{equation}
are the conventional Mie coefficients \cite{Mie books}.

\subsection{Internal field}

For the internal field, the complex amplitudes $\boldsymbol{\mathcal{E}}^{(w)}$ and $\boldsymbol{\mathcal{H}}^{(w)}$ of the positive-frequency parts of 
the electric and magnetic components, respectively, are given in the spherical coordinate system by \cite{Barton88,Barton89} 
\begin{eqnarray}\label{p29}
\mathcal{E}_r^{(w)}&=&\frac{1}{r^2}\sum_{l=1}^{\infty}\sum_{m=-l}^{l}l(l+1)c_{lm}S_l(\bar{n}kr)Y_{lm}(\theta,\varphi),
\nonumber\\
\mathcal{E}_\theta^{(w)}&=&\frac{k}{r}\sum_{l=1}^{\infty}\sum_{m=-l}^{l}\bigg(\bar{n}c_{lm}S'_l(\bar{n}kr)\frac{\partial Y_{lm}(\theta,\varphi)}{\partial\theta}
\nonumber\\&&\mbox{}
-md_{lm}S_l(\bar{n}kr)\frac{Y_{lm}(\theta,\varphi)}{\sin\theta}\bigg),
\nonumber\\
\mathcal{E}_\varphi^{(w)}&=&\frac{k}{r}\sum_{l=1}^{\infty}\sum_{m=-l}^{l}\bigg(im \bar{n}c_{lm}S'_l(\bar{n}kr)\frac{Y_{lm}(\theta,\varphi)}{\sin\theta}
\nonumber\\&&\mbox{}
-id_{lm}S_l(\bar{n}kr)\frac{\partial Y_{lm}(\theta,\varphi)}{\partial\theta}\bigg),
\end{eqnarray}
and
\begin{eqnarray}\label{p30}
\mathcal{H}_r^{(w)}&=&\frac{c\epsilon_0}{r^2}\sum_{l=1}^{\infty}\sum_{m=-l}^{l}l(l+1)d_{lm}S_l(\bar{n}kr)Y_{lm}(\theta,\varphi),
\nonumber\\
\mathcal{H}_\theta^{(w)}&=&\frac{c\epsilon_0 k}{r}\sum_{l=1}^{\infty}\sum_{m=-l}^{l}\bigg(\bar{n}d_{lm}S'_l(\bar{n}kr)\frac{\partial Y_{lm}(\theta,\varphi)}{\partial\theta}
\nonumber\\&&\mbox{}
+m\bar{n}^2c_{lm}S_l(\bar{n}kr)\frac{Y_{lm}(\theta,\varphi)}{\sin\theta}\bigg),
\nonumber\\
\mathcal{H}_\varphi^{(w)}&=&\frac{c\epsilon_0 k}{r}\sum_{l=1}^{\infty}\sum_{m=-l}^{l}\bigg(im \bar{n}d_{lm}S'_l(\bar{n}kr)\frac{Y_{lm}(\theta,\varphi)}{\sin\theta}
\nonumber\\&&\mbox{}
+i\bar{n}^2c_{lm}S_l(\bar{n}kr)\frac{\partial Y_{lm}(\theta,\varphi)}{\partial\theta}\bigg).
\end{eqnarray}
The internal-field coefficients $c_{lm}$ and $d_{lm}$ are determined by the formulae
\begin{equation}\label{p31}
\begin{split}
c_{lm}=\frac{n_2^2\xi_l^{(1)\prime}(n_2ka)S_l(n_2ka)-n_2^2\xi_l^{(1)}(n_2ka)S'_l(n_2ka)}{\bar{n}^2S_l(\bar{n}ka)\xi_l^{(1)\prime}(n_2ka)-n_2\bar{n}S'_l(\bar{n}ka)\xi_l^{(1)}(n_2ka)} A_{lm},\\
d_{lm}=\frac{n_2\xi_l^{(1)\prime}(n_2ka)S_l(n_2ka)-n_2\xi_l^{(1)}(n_2ka)S'_l(n_2ka)}{n_2S_l(\bar{n}ka)\xi_l^{(1)\prime}(n_2ka)-\bar{n}S'_l(\bar{n}ka)\xi_l^{(1)}(n_2ka)} B_{lm}.
\end{split}
\end{equation}
Note that Eqs.~(\ref{p31}) can be rewritten in the form
\begin{equation}\label{p31a}
\begin{split}
c_{lm}=c_{l}^{\mathrm{Mie}} A_{lm},\\
d_{lm}=d_{l}^{\mathrm{Mie}} B_{lm},
\end{split}
\end{equation}
where
\begin{equation}\label{p31b}
\begin{split}
c_{l}^{\mathrm{Mie}}=\frac{n_2^2\xi_l^{(1)\prime}(n_2ka)S_l(n_2ka)-n_2^2\xi_l^{(1)}(n_2ka)S'_l(n_2ka)}{\bar{n}^2S_l(\bar{n}ka)\xi_l^{(1)\prime}(n_2ka)-n_2\bar{n}S'_l(\bar{n}ka)\xi_l^{(1)}(n_2ka)},\\
d_{l}^{\mathrm{Mie}}=\frac{n_2\xi_l^{(1)\prime}(n_2ka)S_l(n_2ka)-n_2\xi_l^{(1)}(n_2ka)S'_l(n_2ka)}{n_2S_l(\bar{n}ka)\xi_l^{(1)\prime}(n_2ka)-\bar{n}S'_l(\bar{n}ka)\xi_l^{(1)}(n_2ka)}
\end{split}
\end{equation}
are the conventional Mie coefficients \cite{Mie books}.

\end{document}